\documentclass[11pt, superscriptaddress]{revtex4-2}
\usepackage{graphics,epsfig}
\usepackage{epsfig}
\usepackage{graphicx}
\usepackage{dcolumn}
\usepackage{amsmath}
\usepackage{amsfonts}
\usepackage[colorlinks=true, pdfstartview=FitV, linkcolor=blue, citecolor=red, urlcolor=magenta]{hyperref}
\usepackage{color}
\usepackage{xcolor}
\begin{document}
\title{Thermal analysis of black hole in de Rham--Gabadadze--Tolley  massive gravity in Barrow entropy framework}
\author{Muhammad Yasir}
\email{yasirciitsahiwal@gmail.com}\affiliation{Department of Mathematics, Shanghai University  and Newtouch Center for Mathematics of Shanghai University,  Shanghai, 200444, P.R. China}
\author{Xia Tiecheng}
\email{xiatc@shu.edu.cn}\affiliation{Department of Mathematics, Shanghai University and Newtouch Center for Mathematics of Shanghai University,  Shanghai, 200444, P.R. China}
\author{Sudhaker Upadhyay\footnote{Visiting Associate, Inter-University Centre for Astronomy and Astrophysics (IUCAA), Pune, Maharashtra 411007, India}}
\email{sudhakerupadhyay@gmail.com}\affiliation{Department of Physics, K. L. S. College, Magadh University, Nawada, Bihar 805110, India}
\affiliation{School of Physics, Damghan University, P.O. Box 3671641167, Damghan, Iran}

\begin{abstract}
This study examines a recently hypothesized black hole solution in  de Rham--Gabadadze--Tolley  massive gravity. Firstly,  we consider the negative cosmological constant as a  thermodynamic pressure. We extract the  thermodynamical properties  such as Hawking temperature, heat capacity  and  Gibbs free energy using the Barrow entropy. We also obtain a new pressure associated  to the  perfect fluid dark matter and discuss the  first-order van der Waals-like phase transition. This black hole's stability is investigated  through specific heat and Gibbs free energy.  {Also, we analyze the thermodynamic curvatures behavior of black hole through geometry methods (Weinhold, Ruppeiner, Hendi-Panahiyah-Eslam-Momennia (HPEM), and geometrothermodynamics (GTD)).}
\end{abstract}

\maketitle
\section{Introduction and Motivation}

One of the most exciting and stimulating area of study is the geometrical structure of black hole (BH) in general theory of relativity (GR) \cite{1}. Exciting phenomena in  de Rham--Gabadadze--Tolley (dRGT) massive gravity testing the recognition of D-bound and Bekenstein bound provides the phantom background and excluding quintessence fields with cosmological constant. In Ref.\cite{m1}, there is discussion of entropy bounds testing the plausibility of the cosmological constant includes other areas or more details, refer to \cite{m2}.

{ Additionally, the Bekenstein limit displays the highest entropy limit on the physical system (Universe-BH system). This change (D bond) does not provide the  guarantee of  physical behavior, so no additional energy can be introduced into the Bekenstein bond. It is worth to  mentioning here that there are  many efforts  to proposing extensions of Einstein's general theory of relativity to explain the current accelerating expansion of the Universe. In GR, the curvature singularity is an identification of breakdown; this is broadly revealed to the existence and its requisite in quantum gravity because it is a more primitive theory of gravitation \cite{t1}.
For the first time, Hawking investigated the BH temperature, in this case he observed that BH can emit radiations near the event horizon from side-to-side gravitational interactions. Also, Hawking and Page constructed  the  stable phase transition between the Schwarzschild BH and the pure radiation (gas) \cite{4,t2}.
The quantum scenario concerning Hawking radiation, this gives us a precious clue that a BH has its temperature directly connected to its area gravity and that its entropy is proportional to the horizon area. These results have shown that there exist a deep association between thermodynamics and gravity \cite{t8}. Later on, the phase transition and liquid-gas phase transition  were investigated for charged and rotating AdS BHs; in this case, there exists a small or large  BHs and van der Waals fluid  \cite{t3,t4,t5}.}
On other hand, in extended phase space thermodynamics studies where we put the key association between the cosmological constant (thermodynamic pressure) and its conjugate variable (thermodynamic volume) \cite{t6}.  Most impressive characteristics studied  for AdS BHs thermodynamics through quantum gravity space-time to have BH horizon, which provides an accumulation of concepts from the GR  and the thermodynamics in quantum field theory\cite{t7}.

{Afterward, the thermodynamic characteristics of BHs have been widely studied, such as thermodynamical behavior for  BHs through  Riemannian geometry \cite{g1,g2,t9}. This Ruppeiner's metric by taking the thermodynamic potentials corresponding to the mass by Legendre transformations \cite{g6,g7}. The construction of thermodynamic geometry in the context of nonlinear electrodynamics, this provides the interpretation of the microstructure for various BHs. It is testing to generalize the  thermodynamic curvature to understand  the microstructure of  BHs; in this scenario, it is indelible complications  relating  to the first law of BH and Lagrangian formations are integrated with nonlinear electrodynamics rather than linear ones. In particular, to study the direct relation between the phase transitions and curvature singularities of Weinhold's \cite{g3} and Ruppeiner's \cite{g4,g5} interpretations, which are manifested through a Hessian matrix in terms of entropy and internal energy respectively. Recently, numerous approaches have been made to inaugurate the different geometries phenomenon  into the ordinary thermodynamics \cite{g1,g2,g3,g4,g5}. Hermann \cite{g8} investigated the effects of the thermal geometries phase space system  as a differential manifold (natural contact structure), which provides  a particular subspace of equilibrium states in the thermodynamics. A new metric Hendi-Panahiyah-Eslam-Momennia (HPEM metric) was introduced in order to build a geometrical phase space by thermodynamical quantities which shows that the characteristics behavioral of Ricci scalar $R^{HPEM}$ of this metric enables one to recognize the type of phase transition and critical behavior of the BHs near phase transition points. Also,  the geometrothermodynamics (GTD) of this BH and investigated the adaptability of the curvature scalar of geothermodynamic methods with phase transition points of the BH. Also, in GTD of this black hole and investigated the adaptability of the curvature scalar of geothermodynamic methods with phase transition points of  BH. The geometrical interpretation is a great deal for exploring the microstructure of BHs thermodynamic system \cite{g6,g7,g8,g9,g10,g11,g12,g13,g14,g15,g16,g17,g18}.   According to one thought, interactions in the system under the study of microscopic statistical features have been illustrated through thermodynamic amounts induced from these  geometries \cite{g11,g12,g13,g14,g15,g16,g17,g18,g19}. Even though the geometrothermodynamics (GTD) methods for various BHs are studied, it remains unexplored in case of BH in dRGT massive gravity via Barrow entropy. This provides us a good opportunity to bridge the gap.}

{In this paper, we use thermodynamic Barrow  entropy and study thermodynamic quantities as well as thermodynamic geometric methods for  BH in massive gravity dRGT. To evaluate the Barrow entropy of the BH in the massive gravity dRGT  due to thermal fluctuations, we exploit the expressions for the Hawking temperature and specific heat. Additionally, using standard thermodynamic relationships, we calculate mass and the heat capacity. For the mass in terms of Barrow entropy reaches a maximum value at $S_B = 4.0$ $(=S_{Bm})$, after that it becomes zero at point $S_B = 9.5$ with the fixed values $\alpha$, $\beta$ , $c$  and  $m_g$.
In case of heat capacity, we find that, the heat capacity remains negative in the range of $0 < S_B < 1$ (unstable phase). However, with values of $S_B \leq 1$, this leads to stable phase transition. We study the  behavior of Gibbs free energy $G$, which represents an equilibrium state at constant pressure. The Gibbs free energy $G$ is a fundamental quantity and provides information about the first-order phase transition of BH in the massive gravity dRGT. Furthermore, we analyze the thermodynamic geometry of such BH. To do this, we plot the thermodynamic quantities and scalar curvature  of the Weinhold, Ruppeiner and GTD methods against Barrow entropy. We find that the singular curvature points are scalars of the Weinhold and GTD methods with Barrow entropy. These geometrical methods, along with others, provide invaluable information about the nature of BH, their formation, evolution and related phenomena. In this way, they form the basis of BH physics and contribute significantly to our understanding of the  most mysterious objects in the universe.}

In section \ref{sec2},  we discuss  the fundamental details of BH solution in dRGT massive gravity. Within section, we study the various thermodynamics of the solution together with the thermal stability of BH   in dRGT massive gravity in the presence of Barrow entropy.
 In section \ref{sec3},  we investigate the thermal geometries of BH  with Barrow entropy.
Finally, in section \ref{sec4}, we provide the discussions and final remarks.

\section{Brief Review of BH solution in dRGT massive gravity}\label{sec2}

Einstein's gravity escorts a massive gravity associated with a scalar field.   Therefore, we added a suitable nonlinear term in the Einstein-Hilbert action  as follows  \cite{b0}
\begin{eqnarray}\label{M1}
S=1/(16\pi)\int d^4 x  \sqrt{-g}\left[\Re+m_g^2 \bigsqcup\left(g, \phi^a\right)\right],
\end{eqnarray}
 { where  $ \bigsqcup$ is a potential, and  $\Re$ is the Ricci scalar for the graviton, which changes the action with the graviton parameter $m_g$ which represents the  graviton mass. The action is expressed in terms of units so that $ G=1$. In four-dimensional spacetime, the  effective potential $\mathcal{U}$ is expressed  as follows}
\begin{eqnarray}\label{M2}
 \bigsqcup\left(g, \phi^2\right)= \bigsqcup_2+\alpha_3  \bigsqcup_3+\alpha_4  \bigsqcup_4,
\end{eqnarray}
where constants $\alpha_3$ and $\alpha_4$ represent the dimensionless free parameters of the theory and the symmetric  potentials along with reliance on metric $g$ and scalar field $\phi^a$.
{The term $\kappa^n$ typically represents a parameter or a coupling constant raised to the power of $n$. The parameter $\kappa$ commonly appears in gravitational theories and often represents the gravitational coupling constant. Its precise value depends on the choice of units and the particular gravitational theory being considered.} These can be written as
\begin{eqnarray}\label{M3}
\begin{gathered}
 \bigsqcup_2 \equiv[\kappa]^2-\left[\kappa^2\right], \\
 \bigsqcup_3 \equiv[\kappa^3-3[\kappa]\left[\kappa^2\right]+2\left[\kappa^3\right], \\
 \bigsqcup_4 \equiv[\kappa]^4-6[\kappa]^2\left[\kappa^2\right]+8[\kappa]\left[\kappa^3\right]+3\left[\kappa^2\right]^2-6\left[\kappa^4\right],
\end{gathered}
\end{eqnarray}
{ where $[\kappa]$ is eigenvalues of matrix $\delta^\mu_\nu - \sqrt{g^{\mu\eta}f_{\xi \sigma} \partial_\eta \phi^\xi \partial_\nu \phi^\sigma}$. Where, $f_{\xi \sigma}$ is a reference metric and $[\mathcal{K}]=\mathcal{K}_\mu^\mu$ is a trace. One can split  the gauge condition by selecting the unitary gauge and imposing an additional restriction on the St\"uckelberg scalars. The scalar field $\phi^a$ has an upper index, leading to a convention that can be used to distinguish these fields from other types of variables in the theory. In this case, the use of capital letters for scalar fields can be treated as emphasize to their meaning or conform to symbols commonly used in related fields of the theoretical physics. This is simply a matter of convention for clarity and consistency in the specific context of dRGT massive gravity. We choose a unitary gauge $\phi^a=x^\mu \delta_\mu^a \mid$, in this measurement, five graviton degrees are described by the metric $g_{\mu \nu}$.  St\"uckelberg scalar transformation equals to the  coordinate transformation. According to this selection, it is possible to break the state of measure and make  the additional modifications in the St\"uckelberg scalar. $\phi^a$ is introduced as a scalar field to recover the general covariance so-called St\"uckelberg scalar theory.
The term $\kappa^n$ typically represents a parameter or a coupling constant raised to the power of $n$. The parameter $\kappa$ commonly appears in gravitational theories and often represents the gravitational coupling constant. Its precise value depends on the choice of units and the particular gravitational theory being considered.} For simplicity, one can rewrite the free parameters (constants) $\alpha_3$ and $\alpha_4$  in terms of new parameters $\alpha$ and $\beta$ as
$\alpha_{3}=\frac{\alpha-1}{3}$
and
$\alpha_{4}=\frac{\beta}{4}+\frac{1-\alpha}{12}$.
The static and spherically symmetric BH solutions of the modified Einstein equations can be written  as a line element:
\begin{eqnarray}\label{M4}
d s^2=-f(r) d t^2 +\frac{d r^2}{f(r)}+h(r)^2 d \Omega^2,
\end{eqnarray}
This can be categorized  and arrange  into $d(r)=0$( $h(r)=h_0 r$) here  $h_0$ is a constant in the form of $\alpha$ and $\beta$ respectively \cite{b1,b2,b3,b4}.  One can consider $h(r)=r$ in the line element to have a BH solution. So, a particular class of results of dRGT massive gravity with the symmetries.
{ Therefore, we are interested in these symmetries because this is a good deal in the degree of freedom of the solutions and has an important consequences to discuss the stability. The metric function  $f(r)$ follows as}
\begin{eqnarray}\label{M6}
f(r)=  1-\frac{2 M}{r}+\frac{\Lambda}{3} r^2+\gamma r+\zeta,
\end{eqnarray}
where
\begin{eqnarray}\label{M7}
\begin{gathered}
\Lambda=3 m_g^2(1+\alpha+\beta), \\
\gamma=-c m_g^2(1+2 \alpha+3 \beta), \\
\zeta=c^2 m_g^2(\alpha+3 \beta).
\end{gathered}
\end{eqnarray}
Here, $M$  represents the  integration constant that associated with the BH mass, and $c$  is a constant used in the reference metric. Here the cosmological constant $\Lambda$ depends (linked) to the  graviton mass $m_g$. So, the graviton mass $m_g$ has a connotation for the accelerated Universe expansion. 
{One would be interested in whether the BH solutions in this theory can converge to the Schwarzschild solution in general relativity under certain conditions. Research in this area typically includes to studying the equations of motion and solutions extracting from the dRGT theory and analyzing whether they reduce to the well-known Schwarzschild solution in general relativity with appropriate limits, such as when the graviton mass $m_g$ and free parameters $(\alpha, \beta)$ become zero $m_g,\alpha, \beta=0$.}

\subsection{Barrow Entropy}
In Ref.  \cite{15a}, Barrow studied the prospects  that quantum-gravitational forces could provide an impenetrable, modifying the BH real horizon region and  fractal structure on the BH surface.  { According to Barrow, the effects of quantum gravity change the standard entropy relation  $S=\frac{A}{4}$ to $S=A^{1+\frac{\Delta}{2} }$, where $A $ is the actual surface and $\Delta$ represents the quantum gravitational distortion effect. Keeping in mind the importance of  Barrow  entropy, we study the impact of  BH thermodynamics in dRGT theory. We analyze the physical engine of the Barrow formula. The Bekenstein-Hawking entropy plays a fundamental role in the three-dimensional model of dark energy, where $S_{BH} = A/(4G)$ and it is used at the \cite{R1,R2} horizon. Although Barrow BH entropy was built as a toy model, some theoretical evidence supports its idea.} As a result, a new BH entropy relation follows as
\begin{eqnarray}\label{M8}
S_{B}=\bigg(\frac{A}{A_{0}}\bigg)^{1+\frac{\Delta}{2}}.
\end{eqnarray}
Here, $A_{0}$  represents the Planck area; at this stage, $A$ is the ordinary horizon area. Besides, it is linked with  Tsallis entropy (nonextensive) expression \cite{16a}; this  entropy is different from the  quantum corrected entropy \cite{17a}  with logarithmic. It provides the most straightforward construction for distinctive values such as  $\Delta=0$. One can obtain the well-known result as  Bekenstein-Hawking entropy with  $\Delta=0$; we have the maximal deformation in this case.
\subsection{Thermal Stability}
\begin{figure}
\begin{minipage}{14pc}
\includegraphics[width=16pc]{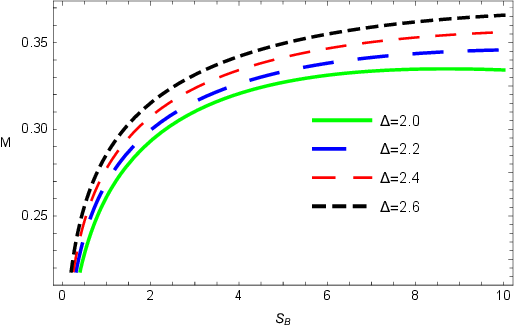}
\caption{\label{1} Plot of mass vs the Barrow entropy with fixed parameters as $\alpha =0.30$, $\beta =0.20$, $c=2.50$ and $m_g=0.25$.}
\end{minipage}\hspace{3pc}%
\begin{minipage}{14pc}
\includegraphics[width=16pc]{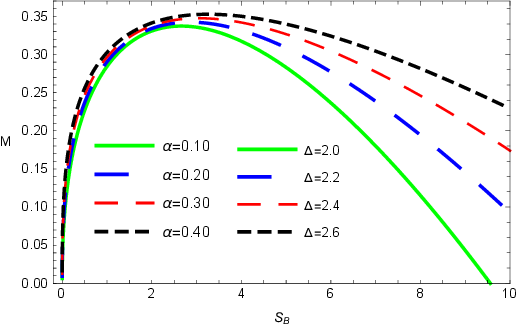}
\caption{\label{2} Plot of mass vs the Barrow entropy with fixed parameters as  $\beta =0.20$, $c=2.50$ and $m_g=0.25$.}
\end{minipage}\hspace{3pc}%
\end{figure}

\begin{figure}
\begin{minipage}{14pc}
\includegraphics[width=16pc]{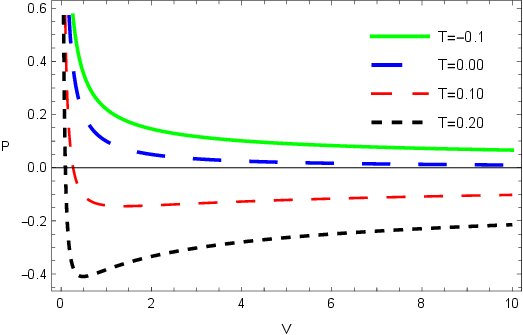}
\caption{\label{f3} { Plot of pressure  $P$ vs. $V$ with fixed  parameters such as $M=0.3$, $\alpha =0.5$,  $\beta =0.02$, $c=1$ and $m_g=0.02$.}}
\end{minipage}\hspace{3pc}%
\begin{minipage}{14pc}
\includegraphics[width=16pc]{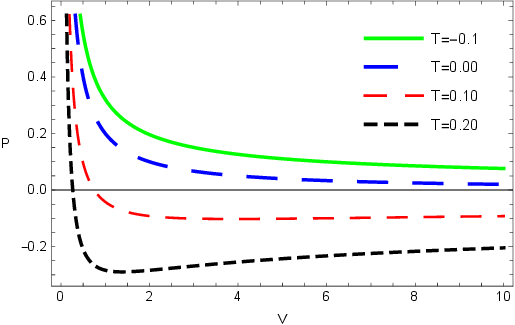}
\caption{\label{f4} { Plot of pressure  $P$ vs. $V$ with fixed  parameters such as  $M=0.6$, $\alpha =0.5$,  $\beta =0.02$, $c=1$ and $m_g=0.02$.}}
\end{minipage}\hspace{3pc}%
\end{figure}

\begin{figure}
\begin{minipage}{14pc}
\includegraphics[width=16pc]{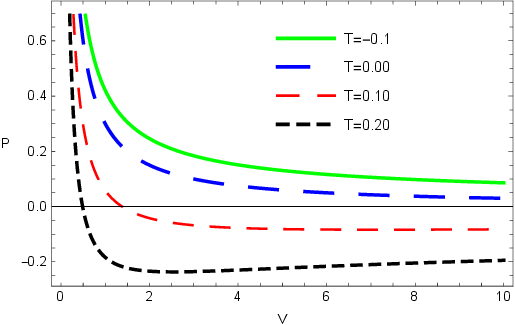}
\caption{\label{f5} { Plot of pressure  $P$ vs. $V$ with fixed  parameters such as   $\alpha =0.5$,  $M=0.9$, $\beta =0.02$, $c=1$ and $m_g=0.02$.}}
\end{minipage}\hspace{3pc}%
\begin{minipage}{14pc}
\includegraphics[width=16pc]{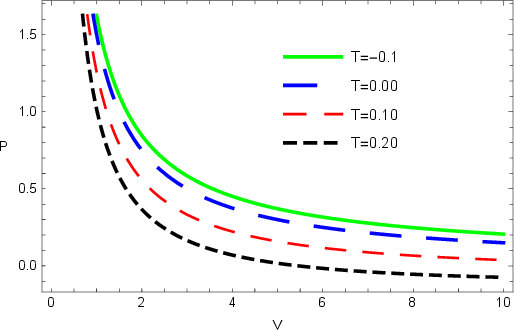}
\caption{\label{f6} { Plot of pressure  $P$ vs. $V$ with fixed  parameters such as  $\alpha =0.5$,  $M=1.2$, $\beta =0.02$, $c=1$ and $m_g=0.02$.}}
\end{minipage}\hspace{3pc}%
\end{figure}

\begin{figure}
\begin{minipage}{14pc}
\includegraphics[width=16pc]{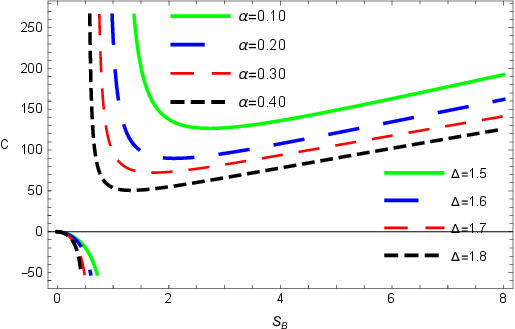}
\caption{\label{f7} Plot of  $C$ with fixed parameters as $\beta =0.20$, $c=0.44$ and $m_g=1.80$.}
\end{minipage}\hspace{3pc}%
\begin{minipage}{14pc}
\includegraphics[width=16pc]{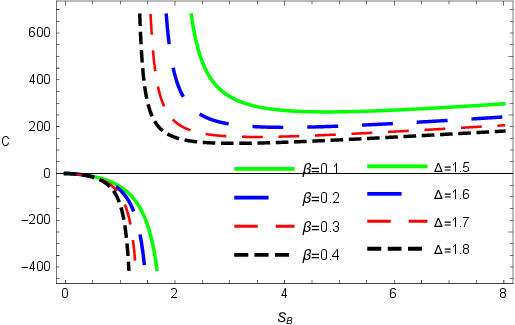}
\caption{\label{f8} Plot of $C$ with fixed parameters as $\alpha =0.20$, $c=0.44$ and $m_g=1.80$.}
\end{minipage}\hspace{3pc}%
\end{figure}

\begin{figure}
\begin{minipage}{14pc}
\includegraphics[width=16pc]{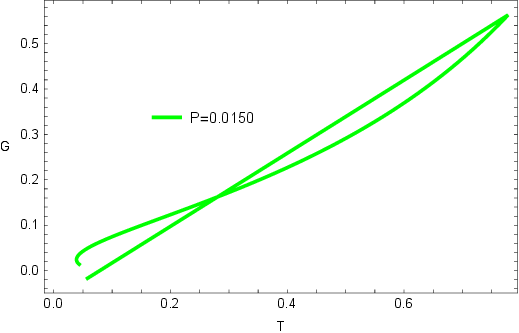}
\caption{\label{f9} Plot of Gibbs free energy  $G$ with fixed parameters as $\alpha =0.10$,  $\beta =0.50$, $c=1.50$, $\Delta =1.40$ and $m_g=1.80$.}
\end{minipage}\hspace{3pc}%
\begin{minipage}{14pc}
\includegraphics[width=16pc]{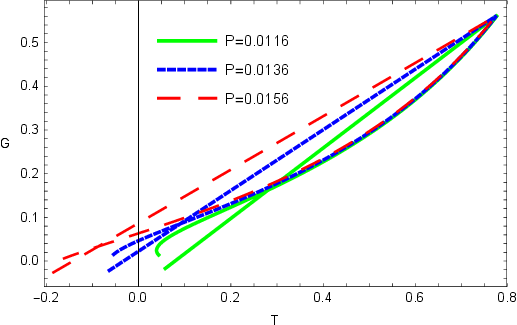}
\caption{\label{f10} Plot of Gibbs free energy  $G$ with fixed parameters as $\alpha =0.10$,  $\beta =0.50$, $c=1.50$, $\Delta =1.40$ and $m_g=1.80$.}
\end{minipage}\hspace{3pc}%
\end{figure}

{In thermal analysis, it is well-known that a cosmological constant  can be considered as a thermodynamic pressure and  the thermodynamic pressure $P$ of the BH can be incorporated into the  first law of thermodynamics. Belonging to the horizon condition $f(r)|_{r=r_h}=0$ along with Barrow entropy and the   pressure relation is $P=-\frac{\Lambda}{8\pi} = \frac{3}{8\pi l^2}$ \cite{t1,t2,t3}, one can extract the expression connecting to the BH mass as follows}
\begin{eqnarray}\label{T1}
M&=& \frac{1}{6}  (\pi ^{-\frac{\Delta }{2}-1} {S_B} )^{\frac{1}{\Delta +2}} \left[3 c^2 m_g^2 (\alpha +3 \beta )-3 c m_g^2 (2 \alpha +3 \beta +1)  (\pi ^{-\frac{\Delta }{2}-1} {S_B} )^{\frac{1}{\Delta +2}}\right.\nonumber\\
&-&\left. 8 \pi  P  (\pi ^{-\frac{\Delta }{2}-1} {S_B} )^{\frac{2}{\Delta +2}}+3\right].
\end{eqnarray}
In  {Figs. \ref{1} and \ref{2}}, we show the effect of BH mass with the Barrow entropy. We see that  in  {Fig. \ref{2}},  the mass of BH reaches a maximum value at $S_B = 4.0$ $(=S_{Bm})$, after that it becomes zero at point $S_B = 9.5$ with the fixed values $\Delta$ and $\alpha$, respectively.
{ The physical and non-physical characteristics of BH can be discussed by analyzing the Hawking temperature. The positive value of temperature ensures the physical solution while the non-physical solution is given by its negative value. After recognizing the thermodynamic behavior of BH, it is worth to  exploring the phase transition points and the divergency points of heat capacity.} With the well-known relation $T=\frac{f'(r_h)}{4 \pi}$, temperature  of BH can be calculated as
\begin{eqnarray}\label{T2}
T&=& \frac{ (\pi ^{-\frac{\Delta }{2}-1} {S_B} ) }{4 \pi}^{-\frac{1}{\Delta +2}}\left[c^2 m_g^2 (\alpha +3 \beta )-2 c m_g^2 (2 \alpha +3 \beta +1)  (\pi ^{-\frac{\Delta }{2}-1} S_B  )^{\frac{1}{\Delta +2}}\right.\nonumber\\
&-&\left.8 \pi  P (\pi ^{-\frac{\Delta }{2}-1} S_B  )^{\frac{2}{\Delta +2}}+1\right].
\end{eqnarray}
The first law of BH thermodynamics can be expressed as \cite{t1,t2,t3,t4,t5}
\begin{equation} \label{T3}
dM=TdS+VdP+ \Phi d\alpha+ \Psi d \beta+ \aleph d m_{g} + C_1 dc,
\end{equation}
where $M$,  $P$,  $V$, $S$, and $\Phi$ are the mass, pressure, volume,  entropy, and chemical potential, respectively. Here, the $\Psi$, $\aleph$ and $C_1$ are the conjugate variables. The thermodynamic parameters symbolize to the conjugating variables  $\beta$, $m_{g}$ and $c$, respectively. The volume $V$ and chemical potential $\Phi$ of BH are defined as
\begin{equation}\label{T4}
V=\bigg(\frac{\partial M}{\partial P}\bigg)_{S,q},\ \ \  \ \
\Phi=\bigg(\frac{\partial M}{\partial \alpha}\bigg)_{S,P}.
\end{equation}
 The Barrow entropy, with the help of first law thermodynamics, follows as
\begin{eqnarray}\label{T5}
V &=& \frac{1}{3} (-4) \pi  \left(\pi ^{-\frac{\Delta }{2}-1} S_B \right)^{\frac{3}{\Delta +2}}, \\
   \Phi &=& -\frac{1}{2} c m_g^2 \left(\pi ^{-\frac{\Delta }{2}-1} S_B \right)^{\frac{1}{\Delta +2}} \left(2 \left(\pi ^{-\frac{\Delta }{2}-1} S_B \right)^{\frac{1}{\Delta +2}}-c\right),\label{T6}\\
\Psi &=& \frac{1}{2} (-3) c m_g^2 \left(\pi ^{-\frac{\Delta }{2}-1} S_B \right)^{\frac{1}{\Delta +2}} \left(\left(\pi ^{-\frac{\Delta }{2}-1} S_B \right)^{\frac{1}{\Delta +2}}-c\right), \label{T7}\\
\aleph &=& c m_g \left(\pi ^{-\frac{\Delta }{2}-1} S_B \right)^{\frac{1}{\Delta +2}} \left(c (\alpha +3 \beta )-(2 \alpha +3 \beta +1) \left(\pi ^{-\frac{\Delta }{2}-1} S_B \right)^{\frac{1}{\Delta +2}}\right), \label{T8}\\
C_1 &=& -\frac{1}{2} m_g^2 \left(\pi ^{-\frac{\Delta }{2}-1} S_B  \right)^{\frac{1}{\Delta +2}} \left((2 \alpha +3 \beta +1) \left(\pi ^{-\frac{\Delta }{2}-1} S_B \right)^{\frac{1}{\Delta +2}}-2 c (\alpha +3 \beta )\right) \label{T9}.
\end{eqnarray}
The pressure can be expressed as
\begin{eqnarray}\label{T10}
P = \frac{-6 \alpha  c m_g^2 V^{\frac{2}{3} }-9 \beta  c m_g^2 V^{\frac{2}{3}}-3 c m_g^2 V^{2/3}+4 \sqrt[3]{6} \pi ^{\frac{2}{3}} M-12 \pi  T V^{\frac{2}{3}}}{8 \sqrt[3]{6} \pi ^{\frac{2}{3}} V}.
\end{eqnarray}
The effect of Barrow entropy is seen in the progression plots in  {Figs. \ref{f3}-\ref{f6}}, where the pressure consistently proscribes the original behaviour of all the isotherms and squeezes them to the critical isotherm. { The green curve represents the behavior greater  than critical  point $T>T_c$}.  Also, it appears that the Barrow entropy parameter removes the fluctuating isotherms at the upper limit of its strength. This may be presented as the maximum temperature value $T$ that exterminates the van der Waals-like nature of the BH in dRGT massive gravity.
With the help of the standard definition of heat capacity \cite{t1, t9}
\begin{equation}\label{T11}
C= \bigg(\frac{\partial S}{\partial T}\bigg)_{P} T,
\end{equation}
one can be calculated as
\begin{eqnarray}\label{T10}
C=\frac{2 \pi  (\Delta +2) S_B  (\pi ^{-\frac{\Delta }{2}-1} S_B  )^{\frac{3}{\Delta +2}-1}   \ (\pi ^{-\frac{\Delta }{2}-1} S_B  )^{\frac{\Delta}{\Delta +2}} }{-c^2 m_g^2 (\alpha +3 \beta )+c m_g^2 (2 \alpha +3 \beta +1)  (\pi ^{-\frac{\Delta }{2}-1} S_B  )^{\frac{1}{\Delta +2}}-1}.
\end{eqnarray}
 {To evaluate the Barrow entropy of the BH in the massive gravity dRGT  due to thermal fluctuations, we exploit the expressions for the Hawking temperature and specific heat. Additionally, using the standard thermodynamic relationships, we plotted the graph of heat capacity  which provides the first-order phase transition. In case of heat capacity, we obtain that, the heat capacity remains negative in the range of $0 < S_B < 1$ (unstable phase). However, with values of $S_B \leq 1$, this leads to stable phase transition in {Figs.  \ref{f7} and \ref{f8}}. We study the behavior of the Gibbs free energy $G$, which represents  equilibrium  at constant pressure. The Gibbs free energy $G$ is a fundamental quantity and provides information about the first-order phase transition of BH in the massive gravity dRGT.}  However, with the specified values of $S_B \leq 1$, one can observe an unstable to stable region, and the stable phase transition (see after the values of $S_B=1$). Finally, in the presence of Barrow entropy, we present an unstable to a stable phase transition with fixed values $M=1.2$, $\alpha =0.5$,  $\beta =0.02$, $c=1$ and $m_g=0.02$.
The Gibbs free energy   can be estimated from the well-known definition as
\begin{eqnarray}\label{T12}
G=M-T S_B.
\end{eqnarray}
This leads to
\begin{eqnarray}\label{T13}
G=2^{\frac{1}{3} (-2) (\Delta +6)} 3^{\frac{\Delta}{3}} \pi ^{\frac{\Delta -4}{6}} V^{\frac{\Delta}{3}} \left(6^{\frac{2}{3}} c m_g^2 V^{\frac{2}{3}} (2 \alpha +3 \beta +1)-8 \pi ^{\frac{2}{3}} M-16 \pi ^{\frac{2}{3}} P V\right)+M.
\end{eqnarray}
{Let us discuss here the Euclidean method also to calculate the Gibbs free energy. 
In the Euclidean action $\mathcal{S}_E$ is expressed by the analytic continuation of the action $S=iS_E$ \cite{R00,R01,R02,R03}, this is associated to the partition function
\begin{eqnarray}\label{r1}
Z=e^{-\mathcal{S}_E},
\end{eqnarray}
where time is defined as $\tau=it$. According to statistical mechanics, the relationship between  free energy and  partition function is
\begin{eqnarray}\label{r2}
G=-T \ln \mathcal{Z}=T \mathcal{S}_E.
\end{eqnarray}
In addition, to  calculate the expression of temperature, we apply a fixed temperature boundary condition throughout Euclidean time as \cite{R00, R02}
\begin{eqnarray}\label{r3}
\int d \tau=\frac{1}{T \sqrt{f\left(r_B\right)}}.
\end{eqnarray}
On the other hand, the time period  $\tau$ is given by the inverse of the Hawking temperature $1 / T_h$, this can written as 
\begin{eqnarray}\label{r4}
T=\frac{T_h}{\sqrt{f\left(r_B\right)}}.
\end{eqnarray}
The concrete expression for the temperature of a static and  BH solution  in de Rham--Gabadadze--Tolley  massive gravity follows as
\begin{eqnarray}\label{r5}
T=\frac{-\frac{3 r_h^2}{l^2}+ \xi +2 \gamma  r_h+1}{4 \pi   \sqrt{f(r_B)} r_h},
\end{eqnarray}
where
\begin{eqnarray}\label{r6}
f(r_B)=\frac{\frac{r_h^3-r_B^3}{l^2}+(r_B-r_h) (\xi +\gamma  r_B+\gamma  r_h+1)}{r_B}.
\end{eqnarray}
For the static symmetrically metric, we evaluate the Euclidean action related to the metric function as
\begin{eqnarray}\label{r6A}
\begin{aligned}
\mathcal{S}_E= & -\frac{1}{T \sqrt{f\left(r_B\right)}} \int_{r_{h}}^{r_B} \left(-2 f^{\prime}(r)-r f^{\prime \prime}(r)+\frac{2 r}{l^2}\right)dr\\
& -\frac{2}{T \sqrt{f\left(r_B\right)}}\left(f\left(r_B\right)+\frac{r_B^2}{l^2}-\frac{r_B \sqrt{f\left(r_B\right)}}{l}\right), \\
\end{aligned}
\end{eqnarray}
calculated as
\begin{eqnarray}\label{r7}
\begin{aligned}
\mathcal{S}_E=-\frac{2 \left(f(r_B) l^2-\sqrt{f(r_B)} l r_B-\gamma  l^2 r_B+r_h \left(\gamma  l^2-2 r_h\right)+3 r_B^2\right)}{\sqrt{f (r_B)} l^2 T}.
\end{aligned}
\end{eqnarray}
Thus, from Eqs.(\ref{r2}) and (\ref{r7}), the  free energy is expressed as
\begin{eqnarray} \label{r8}
G =\frac{2 \left(-f(r_B) l^2+\sqrt{f(r_B)} l r_B+\gamma  l^2 r_B+r_h \left(2 r_h-\gamma  l^2\right)-3 r_B^2\right)}{\sqrt{f(r_B)} l^2}.
\end{eqnarray}}
In Figs. \ref{f9} and \ref{f10}, we can observe a minimum behavior of $G$, which represents an equilibrium state at constant pressure. In this scenario,
a $G$ in dRGT massive gravity is a fundamental quantity providing phase transitions either small or large BHs. So, we have observed the first-order phase transition. In addition, Gibbs free energy is a  analysis  to provide overall stability of a BH.

{ It is well known that thermodynamics of a black hole include three famous mass formulas, i.e., the first law of thermodynamics, the Bekenstein-Smarr mass formula, and the Christodoulou-Ruffini squared-mass formula. Here, we have only considered the first law of thermodynamics. However, it can also be tested through the the Bekenstein-Smarr relation and the Christodoulou-Ruffini-type squared-mass formula  \cite{00,01,02,03,04}.}
\section{Thermal Geometries}\label{sec3}
{ We study the geometrical structures of the BH surrounded by quintessence field by calculating the Weinhold, Ruppiner, GTD and HPEM curvature scalars. However,  we observed that the positive (stable) BH behaviour for specific range such as $S_B\leq 0.78$. Also, these  trajectories provide the  attractiveness along with the repulsive behaviour of BH. Also for the HPEM formalism, there is another divergency of HPEM metric which coincides with the singular points of heat capacity and therefore we can extract more information rather than other mentioned metrics \cite{g3,g4,g5,g6,g7,g8,g9,g10,g11}}. The scalar curvature can be associated with the relationship  between divergence at the critical point and volume of the system. Generally,  the line element for the BH  is given by
\begin{equation}\label{s1}
d s^2=\frac{S M_S}{\left(\prod_{i=2}^n \frac{\partial^2 M}{\partial \xi_i^2}\right)^3}\left[\Sigma_{i=2}^n\left(\frac{\partial^2 M}{\partial \xi_i^2}\right) d \xi_i^2-M_{S S} d S^2\right].
\end{equation}
where $\xi_i \neq S$ and $M_S$  denotes the derivative of mass $M$ with respect  to the entropy $S$. The intrinsic expression of the metric space follows as
\begin{equation}\label{s2}
g=\left(E^c \frac{\partial \phi}{\partial E^c}\right)\left(\eta_{a b} \delta^{b c} \frac{\partial^2 \phi}{\partial E^c \partial E^d} d E^a d E^d\right),
\end{equation}
where
$\frac{\partial \phi}{\partial E^c}=\delta_{b c} I^b$, $\phi$, $I^b, E^a$ and  in turn,  notions the extensive, thermodynamic potential, and  intensive thermodynamic parameters respectively. This section  is studies for thermodynamic geometries such as Weinhold, Ruppeiner, HPEM and GTD formalisms. Despite numerous works on BH thermodynamics, a details  discussion regarding the  microstructure of BH was still missing. Illustration of Weinhold geometry formulated as
\begin{equation}\label{s3}
g_{ij}^{W}=\partial_i \partial_j M(S_B,l).
\end{equation}
The Weinhold metric follows as
\begin{equation}\label{s4}
ds^2_W=M_{S_B S_B}dS_B^2+M_{l l}dl^2+2 M_{S_B l}dS_B dl,
\end{equation}
with the following matrix form:
\begin{equation}\label{s5}
\begin{pmatrix}\nonumber
M_{S_B S_B} & M_{S_B l}\\ \nonumber
M_{l S_B} & M_{l l} \\
\end{pmatrix}.
\end{equation}
One can obtain the the curvature scalar $(R^W)$ using the above equations, calculated as
\begin{eqnarray}   \nonumber
R^W&=&\left[(\Delta -2) l^2 S_B^{\frac{3}{\Delta +2}} (S_B^{\frac{3}{\Delta +2}} (\pi^{7/2} (\beta +\frac{1}{3}) c l^2 m_g^2 S_B^{\frac{2}{\Delta +2}} (\Delta  (S_B-\frac{3}{2})+2 S_B-\frac{3}{2})-\frac{3}{2} \pi ^3 (\Delta -2) S_B^{\frac{3}{\Delta +2}})\right.\\  \nonumber
&+&\left.\left. S_B^{\frac{1-\Delta }{\Delta +2}+1} (\pi ^{7/2} (\beta +\frac{1}{3}) c (\Delta -2) l^2 m_g^2 S_B^{\frac{2}{\Delta +2}}+\pi ^3 (\Delta +2) S_B^{\frac{3}{\Delta +2}}))\right]\right/\left[12 \pi ^{3/2} S_B (S_B^{\frac{2 (1-\Delta )}{\Delta +2}+1}\right.\\ \label{s6}
&-&\left.\frac{1}{2} (\Delta -2) S_B^{\frac{3}{\Delta +2}} (\sqrt{\pi } (\beta +\frac{1}{3}) c l^2 m_g^2 S_B^{\frac{2}{\Delta +2}}+S_B^{\frac{3}{\Delta +2}}))^2\right].
\end{eqnarray}
 {In this case, we demonstrate the positive and negative behavior of $(R^W)$.  We also discuss the curvature behavior and compare it with the heat capacity shown in Figs. 11 and 12. However, we observed that the positive behavior before the singular point such as  $S_B\leq 0.78$, and the negative after the singular point respectively. Moreover, these trajectories provide the attractiveness along with the repulsive behavior of BH  in de Rham--Gabadadze--Tolley massive gravity. This provides the phenomenological information about either repulsive or attractive characteristics between its molecules}. Also, these  trajectories provide the  attractiveness along with the repulsive behaviour of BH. We investigated that this BH provides phenomenologically information about either repulsive or attractive characteristics between its molecules.
\begin{figure}
\begin{minipage}{14pc}
\includegraphics[width=16pc]{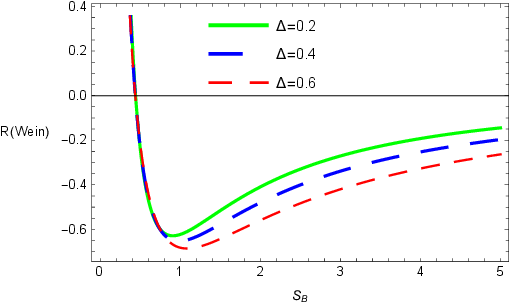}
\caption{\label{f11} Plot of Weinhold curvature scalar $R^W$ with fixed parameters as $\alpha =0.10$,  $\beta =1.0$, $c=0.20$, $l=1.50$ and $m_g=1.10$.}
\end{minipage}\hspace{3pc}%
\begin{minipage}{14pc}
\includegraphics[width=16pc]{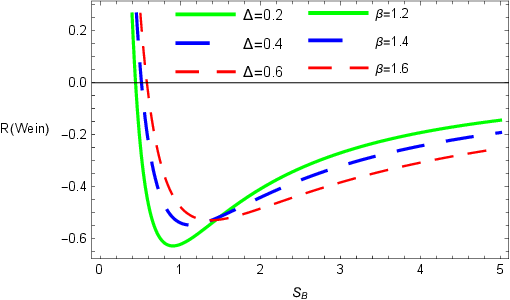}
\caption{\label{f12} Plot of Weinhold curvature scalar $R^W$ with fixed parameters as $\alpha =0.10$, $c=0.20$, $l=1.50$ and $m_g=1.10$.}
\end{minipage}\hspace{3pc}%
\end{figure}
The Ruppeiner geometry conformal (corresponding) to the Weinhold geometry, scalar curvature $R^{Rup}$ is determined as
\begin{equation}\label{s7}
ds^{2}_{R}=\frac{ds^{2}_{W}}{T}.
\end{equation}
The explicit expression of Ruppeiner geometry is derived as
\begin{eqnarray} \nonumber
R^{Rup}&=& \left[\pi ^{\frac{5}{2}} (\Delta -2) l^2 S_B^{\frac{8}{\Delta +2}-1} (\sqrt{\pi} (3 \beta +1) c l^2 m_g^2 (-\Delta +2 (\Delta +2) S_B-7)-3 (\Delta -10) S_B^{\frac{1}{\Delta +2}})\right]\times\nonumber\\
&&\left[18 (S_B^{\frac{4-\Delta }{\Delta +2}}
-\frac{1}{6} (\Delta -2) S_B^{\frac{5}{\Delta +2}} (\sqrt{\pi } (3 \beta +1) c l^2 m_g^2+3 S_B^{\frac{1}{\Delta +2}}))^2 (-(3 \beta +1) c m_g^2-\frac{2 (  S_B)}{\sqrt{\pi}l^2}^{\frac{1}{\Delta +2}}\right.\nonumber\\
&+&\left.(\pi ^{-\frac{\Delta }{2}-1} S_B)^{-\frac{2}{\Delta +2}} (m_g^2 (\alpha +\beta +1) (\pi ^{-\frac{\Delta }{2}-1} S_B)^{\frac{3}{\Delta +2}}+(\pi ^{-\frac{\Delta }{2}-1} S_B)^{\frac{1}{\Delta +2}}))\right]^{-1}.
\end{eqnarray}
In this case, the negative values gives imaginary roots. So, we start from zero points are $S_B = 0$ and $S_B = S_{Bm}$, respectively. Also it  matches with zero point of the temperature  as well as  the heat capacity, this represents the phase transition point. The curvature scalar $R^{Rup}$ of BH concerning to the horizon's radius $S_B$ is studied in {Figs.  \ref{f13} and \ref{f14}}.  In the scenario, the singular points concur with the zero points of heat capacity.
\begin{figure}
\begin{minipage}{14pc}
\includegraphics[width=16pc]{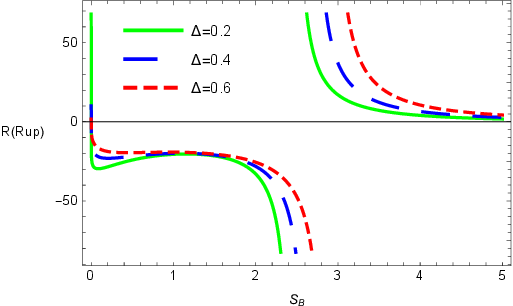}
\caption{\label{f13} Plot of Ruppeiner curvature scalar $R^{Rup}$ with fixed parameters as $\alpha =0.10$,  $\beta =1.80$, $c=0.20$, $l=1.50$ and $m_g=0.07$.}
\end{minipage}\hspace{3pc}%
\begin{minipage}{14pc}
\includegraphics[width=16pc]{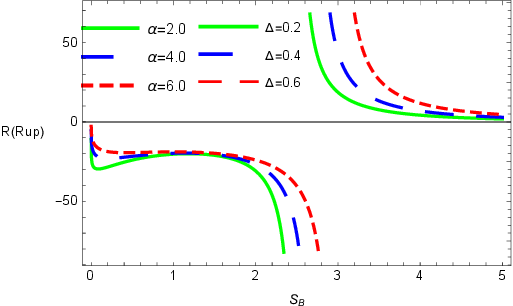}
\caption{\label{f14} Plot of Ruppeiner curvature scalar $R^{Rup}$ with fixed parameters as  $\beta =1.80$, $c=0.20$, $l=1.50$ and $m_g=0.07$.}
\end{minipage}\hspace{3pc}%
\end{figure}
Hence, we observed the thermodynamic phase transition for  curvature scalar of the Ruppiner metric in terms of $S_B$.
{ A new metric HPEM  was introduced in order to build a geometrical phase space by thermodynamical quantities. We show that the characteristics behavioral of Ricci scalar $R^{HPEM}$ of this metric enables one to recognize the type of phase transition and critical behavior of the BHs near phase transition points. This provides us an opportunity to bridge this gap. As a consequence the curvature can be interpreted as a measure of thermodynamic interaction}. This is the motivation of present study. The HPEM geometry follows  as
\begin{equation} \label{s9}
ds^2=\frac{S_B M_{S_B}}{\left(\frac{\partial^2 M}{\partial l^2}\right)^3}(-M_{S_BS_B}dS_B^2+M_{l l}dl^2).
\end{equation}
 The mathematically expression for the HPEM geometry scalar is calculated as
\begin{eqnarray} \nonumber
R^{(HPEM)}&=&\bigg[l^8 S_B^{-\frac{6}{\Delta +2}-1} (3 \sqrt{\pi } \beta  c l^2 m_g^2 S_B^{\frac{2}{\Delta +2}}+\sqrt{\pi } c l^2 m_g^2 S_B^{\frac{2}{\Delta +2}}+3 S_B^{\frac{3}{\Delta +2}})\times\nonumber\\
&& (\pi ^{7/2} (\beta +\frac{1}{3}) c l^2 m_g^2 S^{\frac{2}{\Delta +2}} (\Delta  (S_B-\frac{3}{2})+2 S_B-\frac{3}{2})
-\frac{3}{2} \pi ^3 (\Delta +2) S_B^{\frac{3}{\Delta +2}})\bigg]\nonumber\\  &&\times \bigg[108 (\Delta +2)^2 (\sqrt{\pi } (\beta +\frac{1}{3}) c l^2 m_g^2 S_B^{\frac{2}{\Delta +2}}+S_B^{\frac{3}{\Delta +2}})^2\bigg]^{-1}.\label{s10}
\end{eqnarray}
From the  Figs. \ref{f15} and \ref{f16}, one can see the divergency of scalar curvature of the HPEM  metric at zero point; it is close to the heat capacity plot with fixed parameters  $\beta =1.80$, $c=0.20$, $l=1.50$ and $m_g=0.07$, respectively. Therefore, we extract some beneficial information from the HPEM formalism.
\begin{figure}
\begin{minipage}{14pc}
\includegraphics[width=16pc]{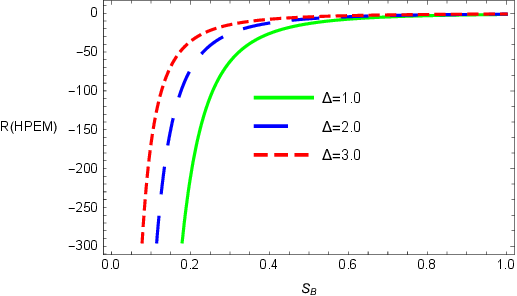}
\caption{\label{f15} Plot of HPEM curvature scalar $R^{(HPEM)}$ with fixed parameters as $\alpha =2$,  $\beta =1.80$, $c=0.20$, $l=1.50$ and $m_g=0.07$.}
\end{minipage}\hspace{3pc}%
\begin{minipage}{14pc}
\includegraphics[width=16pc]{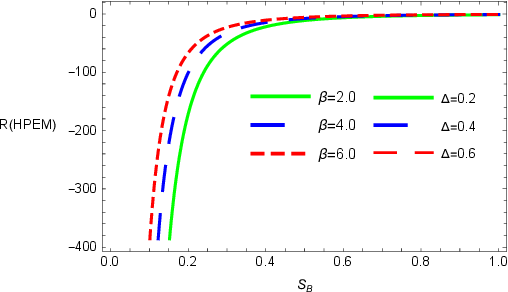}
\caption{\label{f16} Plot of HPEM curvature scalar $R^{(HPEM)}$ with fixed parameters as $\alpha =2$, $c=0.20$, $l=1.50$ and $m_g=0.07$.}
\end{minipage}\hspace{3pc}%
\end{figure}
The scalar curvature for  GTD metric can be formulated as
\begin{equation}\label{s11}
ds^2=(-S_B M_{S_BS_B}+l M_{l l }) \begin{pmatrix}
M_{S_BS_B} & 0\\
0 & m_{l l}\\
\end{pmatrix}.
\end{equation}
From the above expression, one can obtain as
\begin{eqnarray} \nonumber
 R^{(GTD)}&=&  \bigg(\pi ^{3/2} S_B^{-\frac{\Delta +4}{\Delta +2}} (\pi ^{-\frac{\Delta }{2}-1} S_B)^{\frac{2}{\Delta +2}} ((3 \beta +1) c l^2 m_g^2+(2 \Delta +7) (\pi ^{-\frac{\Delta }{2}-1} S_B)^{\frac{1}{\Delta +2}}) \nonumber\\
 &&(9 (\Delta +2) S_B^{\frac{1}{\Delta +2}}
 -\sqrt{\pi } (3 \beta +1) c l^2 m_g^2 (2 (\Delta +2) S_B-3 (\Delta +1)))\bigg)\nonumber\\
 &&\times \bigg[2 (\Delta +2)^2 (\sqrt{\pi} (3 \beta +1) c l^2 m_g^2+3 S_B^{\frac{1}{\Delta +2}})^2\bigg]^{-1}.\label{s12}
\end{eqnarray}
We study the GTD curvature scalar behaviour in  {Figs. \ref{f17} and \ref{f18}}. This can be seen that the singular point at $S_B = 1.21$ for curvature scalar of GTD.  Whenever, we plotted  the heat capacity in  {Figs. \ref{f7} and \ref{f8}} with fixed values of  $\beta =0.20$, $c=0.44$ and $m_g=1.80$ respectively. We observed in that case, the heat capacity was  zero at $S_B =1$. On can see that, if the singular point $S_B =1$ of GTD curvature is not coincident (meet) with the heat capacity. In this scenario, we cannot get any physical information concerning to this frame work from the GTD geometry.
\begin{figure}
\begin{minipage}{14pc}
\includegraphics[width=16pc]{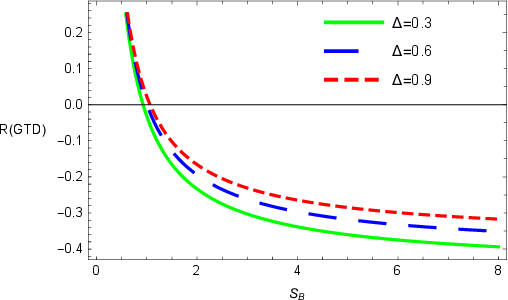}
\caption{\label{f17} Plot of GTD curvature scalar $R^W$ with fixed parameters as $\alpha =2$,  $\beta =1.80$, $c=0.20$, $l=2.50$ and $m_g=0.07$.}
\end{minipage}\hspace{3pc}%
\begin{minipage}{14pc}
\includegraphics[width=16pc]{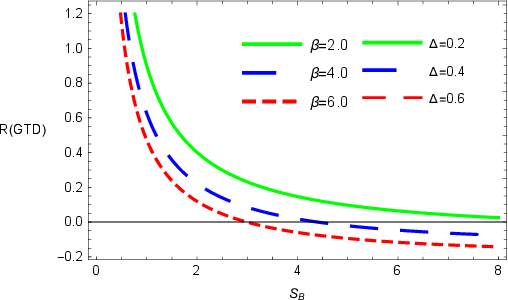}
\caption{\label{f18} Plot of GTD curvature scalar $R^W$ with fixed parameters as $\alpha =2$, $c=0.20$, $l=2.50$ and $m_g=0.07$.}
\end{minipage}\hspace{3pc}%
\end{figure}

\section{Concluding Remarks}\label{sec4}

In this paper, we discussed the thermodynamic behaviour of  BH in dRGT massive gravity and investigated the BH thermal geometries. We graphed thermodynamic properties in terms of $S_B$, and we presented that where it is minimum or  maximum values of the mass, we plotted the mass parameter in  {Figs. \ref{1} and \ref{2}}. The cosmological constant is associated to the state parameter pressure $P$, we studied the phase behaviour in {Figs. \ref{f3}-\ref{f6}}.
In this case, we observed that, if we increase the Barrow entropy parameter, the horizon radius expands and decreases  with critical temperature and pressure .
{ On other hand, two divergences points could be seen  in  {Figs. \ref{f7}-\ref{f10}}. Further, we studied the  first order phase transition (swallow-tail shape)  graphed in  $G - T$. We examined the effect of stability behaviour of heat capacity in dRGT massive gravity using the Barrow entropy.  We investigated the first order  phase transition in the presence of  Barrow entropy, this yields leads to an unstable/stable regions.  Recently,  the similar results have been discussed through thermal quantities of specific BHs  in Ref. \cite{j4, j5,j9,j10}.}

In  {Figs. \ref{f11}-\ref{f14}}, we have observed that the Weinhold and Ruppeiner scalars and matched with singularity points as in heat capacity plots. We found that thermodynamic curvatures such as  HPEM  and GTD have exhibited the attractive as well as repulsive nature of BH for particular parameters. We have investigated the HPEM and GTD geometries of BH in dRGT massive gravity by fixing phase space of entropy, and pressure as shown in  {Figs. \ref{f15}-\ref{f18}}. We also distinguished the behaviour between the specific heat and thermodynamic geometries providing the  beneficial information of such  BH. It would be interesting and stimulating to examine the prospects of similar results also work in most general theories \cite{g12,g18,g19, j7,j8}.

{  Finally, the thermal geometries of  BH in de Rham--Gabadadze--Tolley  massive gravity by using Weinhold, HPEM, Ruppeiner and GTD formalism were presented. We observed the zero of scalar curvature coincided at $S_B = 0.799$ of BH in de Rham--Gabadadze--Tolley massive gravity. Furthermore, important information about the microstructure of BH is provided by Ruppeiner, HPEM and GTD geometries in the form of divergence of scalar curvature, and similar behavior has been observed at zeros of heat capacity in Ref. \cite{R12}. Scalar curvature thus provides important information about the nature of the microscopic interactions described above. These geometric structures thus provide important insight into the critical phenomena and phase structure or divergence of BH in De Rham--Gabadadze-Tolley massive gravity.}

\section*{Acknowledgement}
This project was supported by the Natural Sciences Foundation of China (Grant No. 11975145). The authors thank the reviewers for their comments on this paper.
\section*{Declaration of competing interest}
The authors declare that they have no known competing financial interests or personal relationships that could have appeared to
influence the work reported in this paper.
\section*{Data Availability Statement} This manuscript has no associated data, or the data will not be deposited. There is no observational data related to this article. The necessary calculations and graphic discussion can be made available on request.

\end{document}